# Monolithic photonic integration of suspended light emitting diode, waveguide and photodetector

*Yongjin Wang[1, *], Dan Bai[1], Xumin Gao[1], Wei Cai[1], Yin Xu[1], Jialei Yuan[1], Guixia Zhu[1], Yongchao Yang[1], Xun Cao[2], Hongbo Zhu[1], Peter Grünberg[1]*

[1] Grünberg Research Centre, Nanjing University of Posts and Telecommunications, Nanjing 210003, China

[2] School of Electronic Science and Engineering, Nanjing University, Nanjing, 210093, China

*Corresponding author: wangyj@njupt.edu.cn

*Abstract*

We report here a monolithic photonic integration of light emitting diode (LED) with waveguide and photodetector to build a highly-integrated photonic system to perform functionalities on the GaN-on-silicon platform. Suspended p-n junction InGaN/GaN multiple quantum wells (MQWs) are used for device fabrication. Part of the LED emission is coupled into suspended waveguide and then, the guided light laterally propagates along the waveguide and is finally sensed by the photodetector. Planar optical communication experimentally demonstrates that the proof-of-concept monolithic photonic integration system can achieve the on-chip optical interconnects. This work paves the way towards novel active electro-optical sensing system and planar optical communication in the visible range.

**I. Introduction**

The on-chip optical interconnect concept has been introduced in recent years [1-3], whereby various optical components with different functions are monolithically integrated onto the same chip. On the basis of the nanowire photonics using sophisticated growth technique, compact photonic platforms consisting of light emitting diode (LED) and photodetector, which are connected by waveguide or optically coupled through free space optical transmission, have been demonstrated [4, 5]. To achieve the monolithic integration of emitter, waveguide and detector, it requires a material system with multiple optoelectronic functions simultaneously [6, 7], as well as the similar fabrication process. It is still a challenging issue, especially for further mass production.

GaN-based multiple quantum wells (MQWs) structures grown on silicon substrate are promising and feasible for monolithic photonic integration with the possibility of mass production. With significant advancements in epitaxial growth of high quality GaN-based materials [8, 9], p-n junction GaN-based MQWs can be obtained on various substrates [10, 11]. The GaN-based materials may play the selectable functionalities of light emission, optical transmission and photodetector simultaneously. The p-n junction GaN-based MQWs are excellent emitters and photodetectors, where the aimed wavelength is controllable by tuning the indium or aluminum contents in MQWs [12-14]. GaN is transparent with high optical transmission performance from the visible to the infrared range. By the utilization of large index contrast between GaN and air, the formation of highly-confined structures can be obtained by the removal of silicon substrate and back wafer etching of suspended epitaxial film [15, 16]. On the basis of the p-n junction GaN-based MQWs structures grown on silicon substrate, emitter, waveguide and photodetector are able to be fabricated simultaneously and monolithically integrated into the same epitaxial structure to perform functionalities. Here, we demonstrate a monolithic photonic integration of the LED with the waveguide and the photodetector on the GaN-on-Si platform. To achieve this, suspended p-n junction InGaN/GaN MQWs are used for device fabrication. The investigated monolithic photonic integration can find various applications in planar optical communication, as well as active electro-optical (EO) sensing.

**II. Experimental results and discussion**

The monolithic photonic integration of LED, waveguide and photodetector is implemented on a commercial GaN-on-silicon wafer from Lattice Power Corporation. Schematic diagrams of the fabrication process are shown in Fig. 1. For the p-n junction InGaN/GaN MQWs device, the top layer is firstly etched down to n-GaN to form mesa, as shown in Fig. 1A. The 20nm Ni/ 150nm Au bilayers are used as p- and n-type contacts and simultaneously deposited on the surface of the p-GaN and the n-GaN layers. The device is annealed in an atmosphere of $N_2$ at 500$^o$C for 5 min to obtain ohmic contacts. To separate the LED and the photodetector, part of waveguide is then partially etched to form isolation trench, where the p-GaN and InGaN/GaN MQWs are removed. The processed structures are subsequently protected by thick photoresist, and the silicon substrate underneath the device region is patterned by backside alignment technology. The silicon substrate is removed by deep reactive ion etching, and suspended epitaxial films are thinned by back wafer etching. The monolithic photonic integration of LED, waveguide and photodetector is finally obtained after removing the residual photoresist.

Figure 2A shows the three-dimensional optical micrograph of fabricated monolithic photonic integration of LED, waveguide and photodetector obtained from the backside. The integrated devices are well fabricated on a suspended membrane with 430μm in diameter. Since the GaN is transparent in the visible range, the integrated devices can be clearly observed. Figure 2B demonstrates scanning electron microscope (SEM) image of integrated devices. Because suspended p-n junction InGaN/GaN MQWs membrane device has selectable functionalities either for efficient LED or sensitive photodetector, the LED and the photodetector have the similar structures and are connected by one 80μm long rib waveguide. The gap between mesa and n-electrode is 10μm, and the p-electrode with 60μm in diameter is fabricated on a 70μm×70μm mesa. Three dimensional atomic force microscope (AFM) images of the waveguide and the isolation trench are shown in Fig. 2C. The waveguide has a width of 8μm and the rib height of 620nm, and the isolation trench has a width of 6μm and the depth of 590nm. Suspended membrane illustrated in Fig. 2D has a film thickness of 4.45μm, consisting of ~220nm thick p-GaN layer, ~250nm InGaN/GaN MQWs, ~3.2μm thick n-GaN layer, ~400nm thick undoped GaN layer and ~380nm thick Al(Ga)N buffer layer.

Substrate-free environment is critical for the monolithic photonic integration. If the GaN film remains attached to silicon substrate, the emitted light would simply couple into the silicon substrate and become lost. Moreover, the membrane platform allows suspended LED with advantages of being transferable emitter [17] and with high light extraction efficiencies [18, 19]. The power-dependent electroluminescence (EL) characteristics of suspended membrane LED are measured at room temperature. Figure 3 shows the EL spectra of suspended membrane LED at various applied voltage levels. The emission intensity is enhanced with increasing the applied voltage levels from 4V to 9V, and the dominant EL peak around 454.8nm is stable. The light output intensity of the LED can be modulated at high speed, endowing its potential application for visible light communication [20, 21]. Compared to LED with silicon substrate, the shift of the dominant EL peak takes place owing to the change of stress state. However, the integrated photonic platform could achieve a good spectral matching between the emission and detection since the LED and the photodetector are fabricated on the same suspended membrane.

The light emission images present a direct observation of light coupling among the integrated devices. Figures 4A-4D show the light emission images at different applied voltage levels of 4V to 9V of the LED. These results are in consistent with the measured EL spectra show in Fig. 3. The electrically-driven light emission is power-dependent. The emission intensity is clearly strengthened with increasing the applied voltage levels. The light emission is suppressed in the p-electrode region because of the use of the 20nm Ni/ 150nm Au bilayers as p-contacts. Suspended waveguide allows the light to be channeled making use of total internal reflections along the channel pathway [22-24]. Part of the LED emission laterally couples into the waveguide and propagates along the waveguide. The guided light is then diffracted into the air at the isolation trench facet, as shown in Fig. 4B. With increasing the applied voltage level to 8V and 9V, the isolation trench becomes brighter, and the photodetector is also distinctly observed. These changes reflect the light coupling and lateral propagation in the integrated devices.

The coupling responses among the LED, the waveguide and the photodetector are characterized by four-terminal measurements by a combination of a Cascade Summit 12000M probe station with a Keithley-4200 semiconductor parameter analyzer. Figure 5A shows the induced current at different applied voltage levels of 2V to 9V of the LED. As a p-n junction diode, the photodetector exhibits a rectifying behavior at the applied voltage of 2V. A distinct current increase is observed for bias voltage higher than 1.9V. With increasing the applied voltage, the current-voltage (I-V) curves shift to higher bias voltage levels. The negative currents are clearly observed as the applied voltage is higher than 6V. The maximum negative currents are measured at -0.082μA with the bias voltage of 3.4V and -0.211μA with the bias voltage of 3.7V at the applied voltage levels of 8V and 9V, respectively. To further evaluate the photoresponse of the integrated devices, the devices are illuminated with a 450nm light source, which is generated by a fiber-coupled Ekspla NT200 tunable laser. The power-dependent photoresponses are clearly observed with increasing the illumination power from 0.28mW to 0.42mW, as illustrated in Figs. 5B and 5C. At the applied voltage of 9V, the measured negative currents are -0.236μA with the bias voltage of 3.7V and -0.266μA with the bias voltage of 3.75V at the illumination powers of 0.28mW and 0.42mW, respectively. These results indicate the integrated devices are beneficial to amplify the sensitivity of the photodetector for novel active EO sensors. The transmitted light that interacts with an external object is then sensed by a photodetector to derive information about the object.

Planar optical communication in the visible range is the targeted demonstration of the proof-of-concept monolithic photonic integration. With the LED delivering light signals into the waveguides, the photodetector at the other end of the system will convert the photons back into electrons to complete the information process via optical means. Figure 6A shows the received output signal at the photodetector with the bias voltage of 0V. The LED is directly driven by arbitrary waveform generator to output a 2Mbps random binary sequence. The received signal at photodetector tracks the transmitted signal from the LED, indicating on-chip optical coupling among the LED with the waveguide and the photodetector. Figure 6B illustrates the eye diagram of the monolithic integrated devices. An open eye is clearly shown at 2Mbps. These results demonstrate that the monolithic integrated devices are capable of achieving on-chip optical interconnects for planar visible light communication. Coupling of light from LED into waveguide and from waveguide to photodetector would be improved in future by proper design of the devices. In general, the optical bandwidth of LED is strongly dependent on the injected current density and increased with increasing the current density [25, 26]. The light coupling efficiency among the integrated devices can be improved by optimizing waveguide structures, and the sensitivity of photodetector is able to be enhanced by a proper bias voltage.

### III. Conclusions

In conclusion, we have demonstrated the monolithic photonic integration of suspended p-n junction InGaN/GaN MQWs LED, waveguide and photodetector on a single GaN-on-silicon platform. The LED and the photodetector are connected by a suspended waveguide, and the optical coupling is realized by the in-plane propagation of the emitted light along the waveguide. Both the LED emission and the photodetector sensing are achieved simultaneously. The LED emission is useful to amplify the sensitivity of the photodetector. On the other hand, the data-modulated output of the LED endows the monolithic photonic integration systems with the capability of planar optical communication in the visible range.

**Acknowledgements**

This work is jointly supported by NSFC (11104147, 61322112), research project (2014CB360507, RLD201204, BJ211026).

**Illustration Captions**

**Figure 1.** Schematic illustration of the fabrication processes for monolithic photonic integration of LED, waveguide and photodetector.

**Figure 2.** (a) Three-dimensional optical micrograph of fabricated devices observed from the backside. (b) SEM images of fabricated devices. (c) AFM image of waveguide and isolation trench. (d) SEM image of suspended membrane, and the inset is the zoom-in image of suspended membrane with a film thickness of 4.45μm.

**Figure 3.** Measured EL spectra at different applied voltage levels of 4V to 9V.

**Figure 4.** Optical microscopy images show the light emission at different applied voltage levels of 4 to 9V.

**Figure 5.** (a) I-V curve of suspended photodetector versus the applied voltage levels of the LED. (b) and (c) I-V curve of suspended photodetector versus the illumination powers at the wavelength of 450nm: (b) illumination power of 0.28mW and (c) illumination power of 0.42mW.

**Figure 6.** (a) Transmitted and received signal by a 2Mbps random binary sequence. (b) Eye diagram taken at 2Mbps.

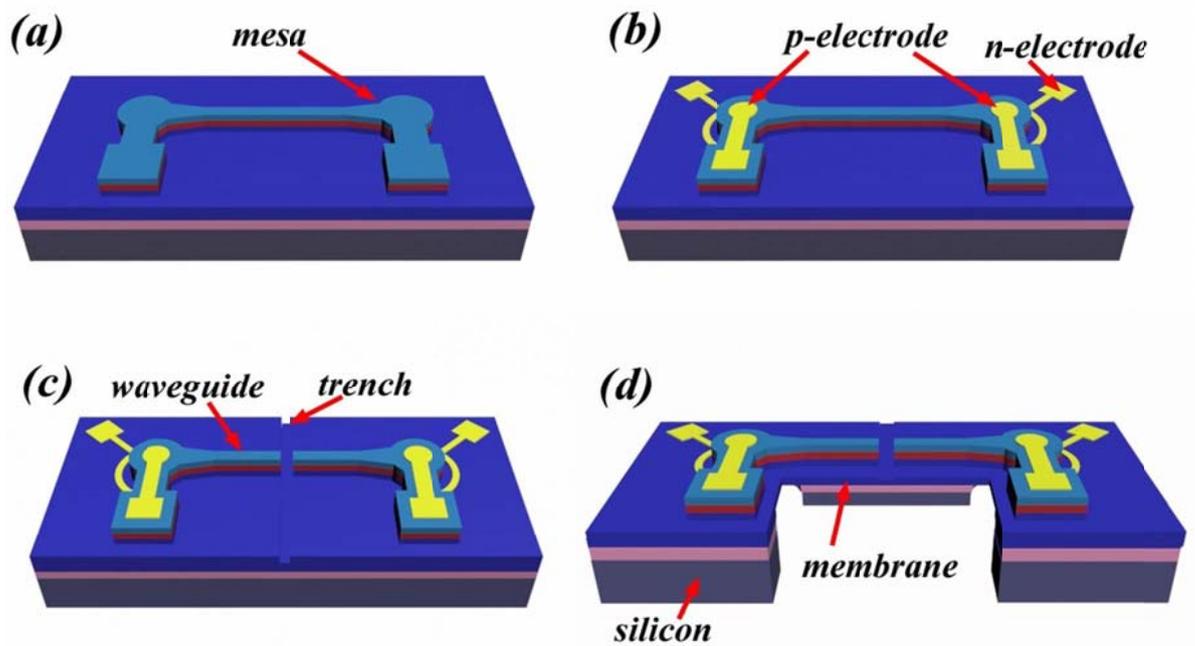

**Figure 1**

*Yongjin Wang et al.*

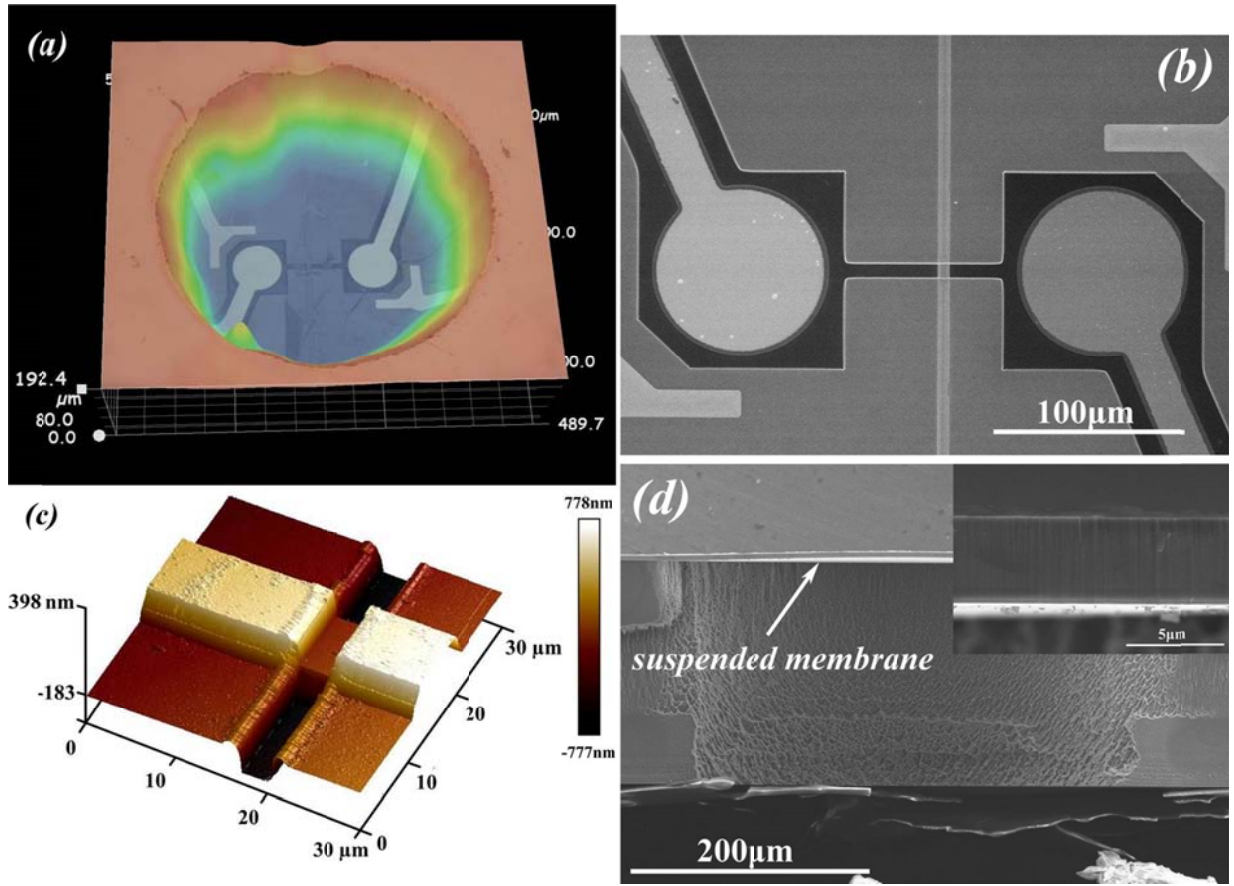

**Figure 2**

*Yongjin Wang et al.*

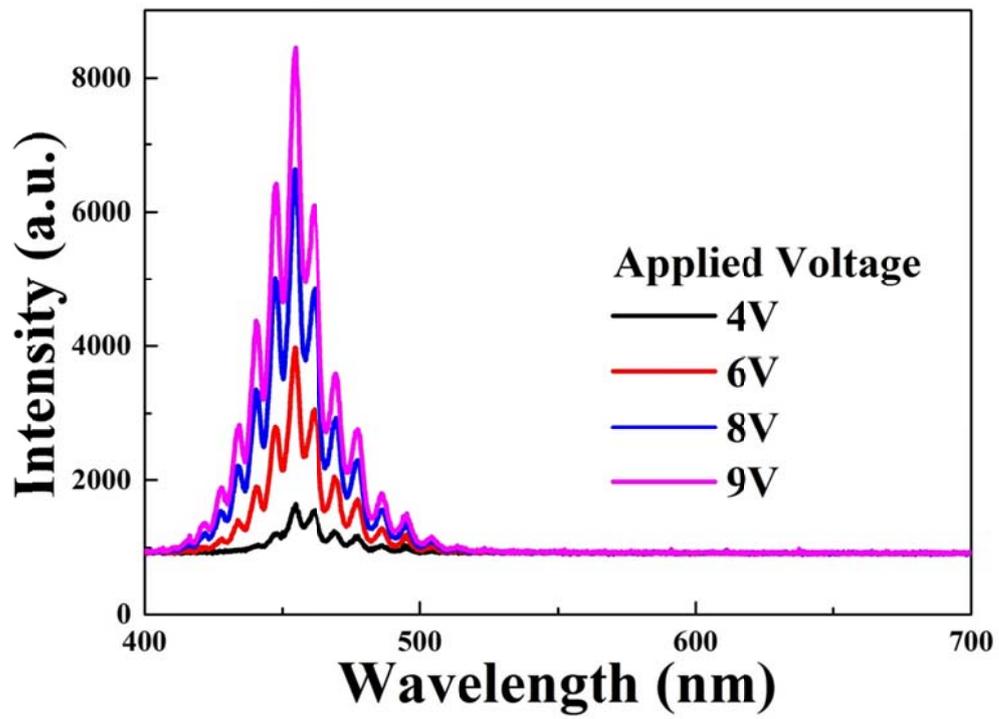

**Figure 3**

*Yongjin Wang et al.*

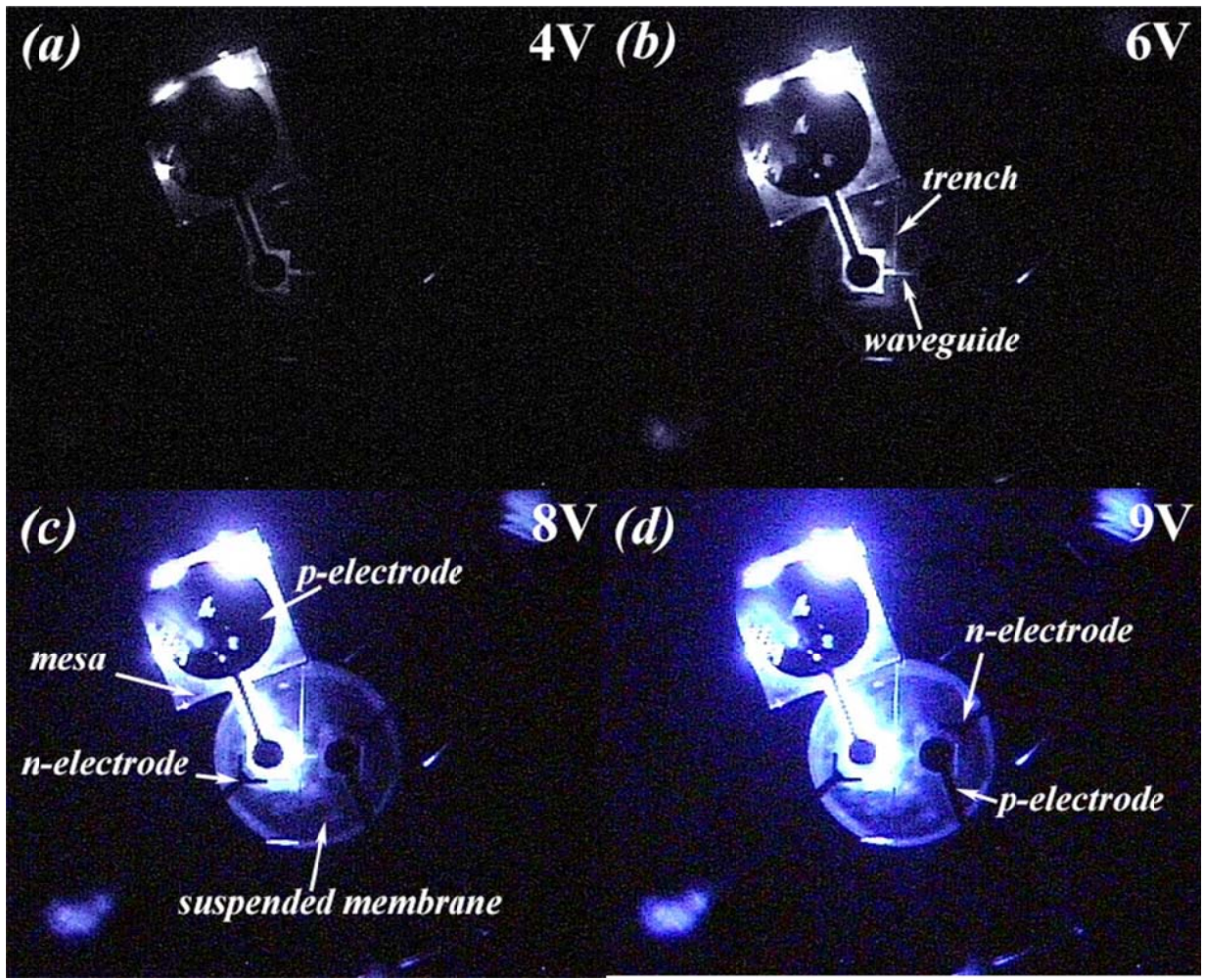

**Figure 4**

*Yongjin Wang et al.*

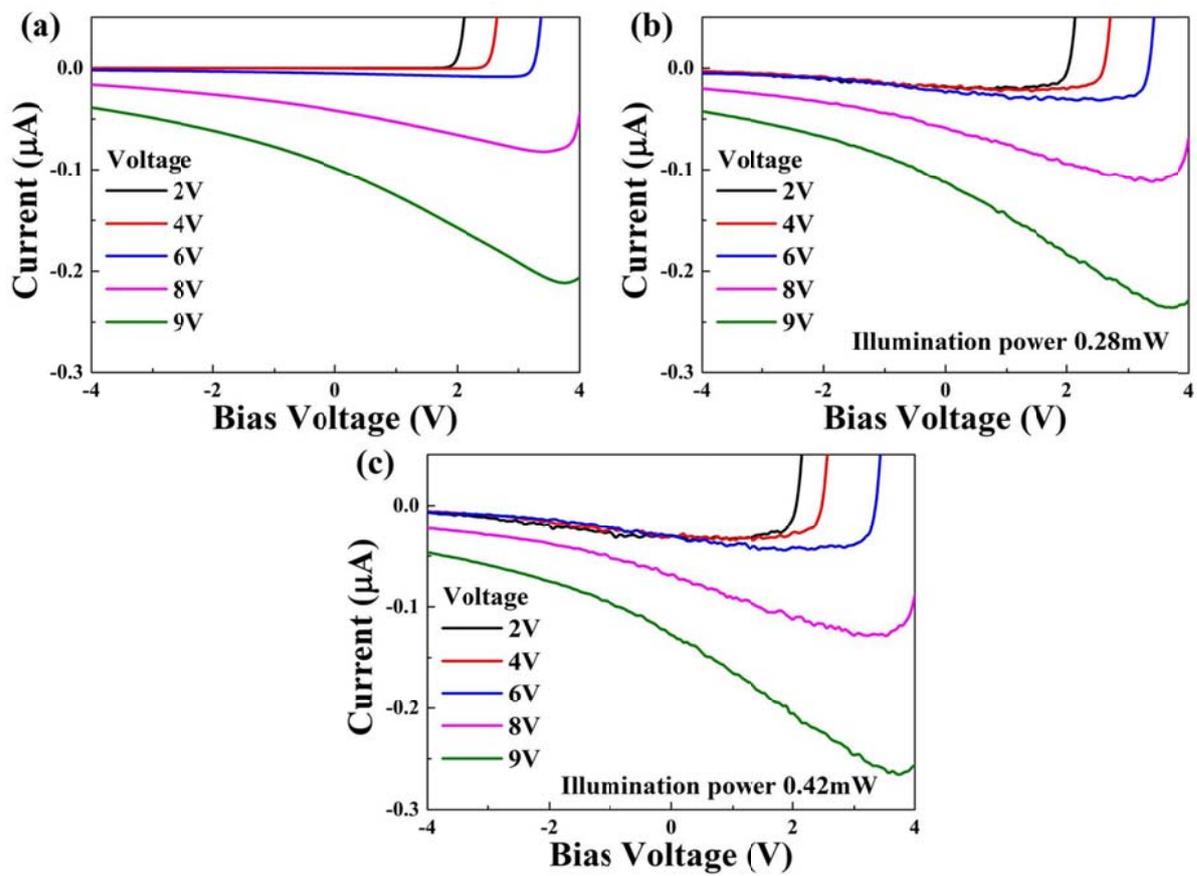

**Figure 5**

*Yongjin Wang et al.*

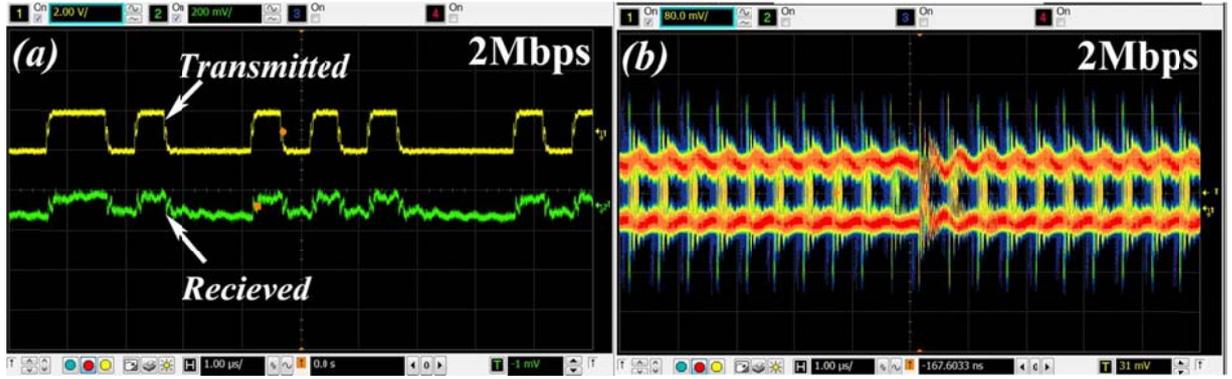

**Figure 6**

*Yongjin Wang et al.*